\documentclass[conference]{IEEEtran}
\IEEEoverridecommandlockouts
\usepackage{amsmath,amssymb,amsfonts,bm}
\usepackage{graphicx}
\usepackage{cite}
\usepackage{booktabs}
\usepackage{array}
\usepackage{multirow}
\usepackage{tabularx}
\graphicspath{{fig/}}
\newcommand{\ul}{\rm{UL}}
\newcommand{\dl}{\rm{DL}}
\newcommand{\fe}{\mathcal{F}^{\rm enc}}
\newcommand{\fd}{\mathcal{F}^{\rm dec}}
\newcommand{\jj}{\mathrm j}
\newcommand{\CN}{\mathcal{CN}} 
\usepackage{xcolor}

\begin{document}
\bstctlcite{bibfbcs:BSTcontrol}
\title{UL-Assisted Parametric CSI Feedback and Prediction for FDD Systems}

\author{%
\IEEEauthorblockN{Hao Chang\IEEEauthorrefmark{1}, Mingcheng Nie\IEEEauthorrefmark{1}, Shuangyang Li\IEEEauthorrefmark{2}, Chang Liu\IEEEauthorrefmark{3}, Branka Vucetic\IEEEauthorrefmark{1},\\ Yonghui Li\IEEEauthorrefmark{1} and Wibowo Hardjawana\IEEEauthorrefmark{1}}
\IEEEauthorblockA{\IEEEauthorrefmark{1}The School of Electrical and Computer Engineering, The University of Sydney, Sydney, Australia\\
\{hao.chang, mingcheng.nie, branka.vucetic, yonghui.li, wibowo.hardjawana\}@sydney.edu.au}
\IEEEauthorblockA{\IEEEauthorrefmark{2}The Department of Electrical Engineering and Computer Science, Technische Universit{\"a}t Berlin, Berlin, Germany, \\ shuangyang.li@tu-berlin.de}
\IEEEauthorblockA{\IEEEauthorrefmark{3}The Department of Computer Science and Information Technology, La Trobe University, Melbourne, Australia, \\ c.liu6@latrobe.edu.au}}
%\vspace{-10mm}
\maketitle

\begin{abstract}
In frequency-division duplex (FDD) systems, the uplink (UL) and downlink (DL) channels are not fully reciprocal. Therefore, DL channel state information (CSI) estimated by each user equipment (UE) must be fed back to the base station (BS). However, because the channel varies between CSI acquisition and utilization, the feedback CSI may become outdated before it is used at the BS. To address this challenge, this paper proposes an UL-assisted parametric CSI feedback and prediction scheme for multi-user FDD systems with time-varying channel parameters. 
By exploiting the partial reciprocity between the UL and DL channel parameters, the BS first estimates the UL channel parameters from the sounding reference signals (SRSs) transmitted by the UEs.
The BS then broadcasts DL pilots, allowing each UE to obtain observations containing information about its current DL channel. Instead of explicitly estimating the DL channel from these observations, each UE directly encodes them into feedback symbols and transmits the symbols to the BS. The BS decoder combines the feedback symbols with the partially reciprocal UL parameters to predict the future DL channel parameters. Finally, the corresponding DL channel matrix is synthesized from the predicted parameters using an explicit physical channel model. Simulation results on Sionna ray tracing (RT) dataset show that the proposed scheme outperforms the considered baselines in terms of channel prediction performance over different signal-to-noise ratios and prediction lengths.

\end{abstract}


\section{Introduction}
Accurate channel state information (CSI) is essential for coherent downlink (DL) transmission \cite{Nie2026ISWCS,11153399,liu2020deepresidual}. In frequency-division duplex (FDD) systems, the uplink (UL) and DL operate at different carrier frequencies, and their channel parameters are not fully reciprocal \cite{9322570}. Therefore, the base station (BS) cannot directly use the UL CSI for DL transmission. Each user equipment (UE) must estimate its DL CSI from the received pilot observations and feed the resulting CSI back to the BS through an UL feedback link \cite{love2008}. The BS then reconstructs the DL CSI for subsequent precoding and data transmission \cite{liu2023predictive}.

Traditional codebook-based CSI feedback follows an estimation, quantization, feedback, and reconstruction chain \cite{jindal2006,love2008}. During DL training, the BS broadcasts pilots to UEs, and each UE estimates its DL CSI from the pilot observations. The CSI estimates are then quantized by searching a predefined codebook for the best-matching codeword, which is fed back to the BS after channel coding and modulation. 
The BS recovers the codeword from feedback information and reconstructs the DL CSI using the same codebook. However, codebook-based CSI feedback schemes may incur substantial feedback overhead and codeword search complexity \cite{9931713}. To alleviate these limitations, neural network (NN)-based CSI feedback schemes have been investigated in recent literature \cite{csinet,GUO202483}.
For instance, CsiNet employs an autoencoder architecture, in which an encoder is used at the UE to compress a DL channel matrix into a low-dimensional representation, and a decoder is then employed at the BS to reconstruct the channel matrix from the received feedback~\cite{csinet}. However, CsiNet assumes that the DL channel matrix is perfectly known at the UE, which is an ideal assumption in practical communication systems. To address this issue, a joint channel estimation and feedback framework was subsequently proposed in~\cite{GUO202483}. It allows the encoder to directly process the received pilot observations, thereby jointly performing channel estimation and feedback.
Nevertheless, the above schemes aim to reconstruct the DL CSI. In practical communication environments, wireless channels exhibit time-varying characteristics due to user mobility and dynamic scattering \cite{11373535,liu2022learning}. Consequently, feedback delay may render the reconstructed DL CSI outdated when it becomes available at the BS.
   
To overcome this limitation, predictive CSI feedback schemes have been developed to predict the DL CSI for a future transmission period rather than reconstructing it \cite{9789200,10720682}.
By exploiting the partial reciprocity between the UL and DL channel parameters, the scheme in \cite{9789200} first estimates the UL channel parameters using the sounding reference signal (SRS). The BS then extracts the angle-delay-Doppler structure from the estimated parameters for the DL precoding. The UE estimates and feeds back a set of complex scalar coefficients, which are used together with the partially reciprocal UL channel parameters to predict the future DL channel at the BS. However, the scheme assumes that the extracted angle-delay-Doppler structure and the related parameters remain nearly unchanged over the prediction interval, which may not hold when the channel parameters vary over time.  
To address this challenge, the Transformer-assisted parametric CSI feedback method in \cite{10720682} predicts the future DL channel matrix under time-varying channel parameters. Specifically, the UE employs a Transformer-based encoder to predict the future DL channel parameters from a sequence of historical DL channel matrices. The predicted parameters are then fed back to the BS, where a Transformer-based decoder outputs the corresponding future DL channel matrix. Nevertheless, processing high-dimensional historical DL channel matrices at the UE and predicting the high-dimensional DL channel matrix at the BS incur substantial computational complexity.  

In this paper, we propose an UL-assisted parametric CSI feedback and prediction scheme that exploits the partial reciprocity between the UL and DL channel parameters in FDD multiple-input multiple-output (MIMO) orthogonal frequency division multiplexing (OFDM) systems. The proposed scheme explicitly accounts for the slow temporal variation of these parameters. 
Specifically, the BS first estimates the UL parameters from the SRSs transmitted by the UEs and then broadcasts pilots to all UEs. At each UE, an encoder directly maps the received pilot observations into feedback symbols without explicit channel estimation or historical DL CSI. 
Compared with compressing a CSI matrix, directly compressing the lower-dimensional pilot observations reduces the encoder complexity. At the BS, a decoder combines the feedback symbols with partially reciprocal UL parameters estimated from SRS observations to predict the low-dimensional parameters of the future DL channel. The high-dimensional DL channel matrix is then synthesized from the predicted parameters using an explicit physical channel model. 
This parametric design exploits the physical channel structure and avoids directly predicting the high-dimensional channel matrix, thereby reducing model complexity.
Simulations on a dataset generated using Sionna ray tracing (RT) \cite{sionnaRT} shows that, under the same feedback budget, the proposed scheme achieves better channel prediction performance than the evaluated CSI feedback baselines over different DL pilot SNRs and prediction lengths, with lower complexity. 

\begin{figure}[!t]
\centering
\includegraphics[width=0.48\textwidth]{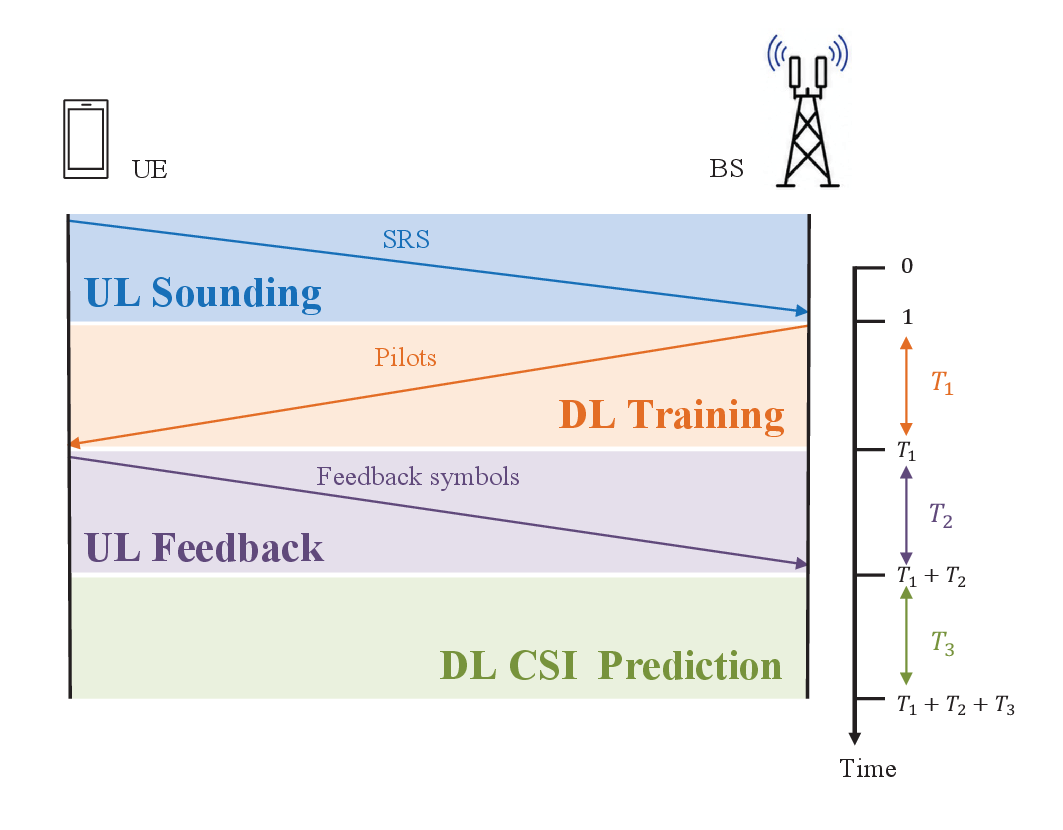}
\vspace{-0.8cm}
\caption{UL and DL transmission phases.}  
\vspace{-0.3cm}
\label{fig:system}
\end{figure}

\textbf{Notation:} Scalars, vectors, and matrices are denoted by $a$, $\mathbf a$, and $\mathbf A$, respectively. The operators $(\cdot)^{\top}$ and $(\cdot)^{H}$ denote transpose and Hermitian transpose. $\|\cdot\|_2$ denotes for $\ell_2$ norm. The operators $\operatorname{Re}\{\cdot\}$ and $\operatorname{Im}\{\cdot\}$  extract the real and imaginary parts, respectively. $\operatorname{vec}(\cdot)$ vectorizes a matrix into a vector. 
$\CN(\mu,\sigma^2)$ denotes complex Gaussian distribution with mean $\mu$ and variance $\sigma^2$. $\otimes$ represents the Kronecker product.

\section{System Model}
Consider a multi-user OFDM system in which a BS is equipped with $M$ antennas and serves $U$ single-antenna UEs over $K$ subcarriers. The UL and DL carrier frequencies are $f_{\ul}$ and $f_{\dl}$, respectively. The subcarrier spacing is $\Delta f$, and $T_{\mathrm s}$ denotes the OFDM symbol duration, with critical sampling $\Delta f \cdot T_{\mathrm s}=1$ \cite{Nie2026SPAWC}.
  
\subsection{Channel Model}  
We consider a stationarity region within which the propagation environment remains approximately unchanged, while the path parameters are slowly time-varying due to UE mobility \cite{10225901,liu2023predictive,nie2024uplink}.
The discrete time-frequency (TF) domain channel for user $u\in\{1,\dots,U\}$, OFDM symbol $t\in\{0,\dots,T\}$, and subcarrier $k\in\{1,\dots,K\}$ is written as

\begin{equation}
\begin{aligned}
\mathbf h^{\rm A}_{u,t,k}
&=\sum_{\ell=1}^{L}\beta^{\rm A}_{u,t,\ell}
\mathbf a^{\rm A}\!\left(\phi^{\rm A}_{u,t,\ell},\theta^{\rm A}_{u,t,\ell},f^{\rm A}\right)\\
&\quad\times
 e^{-\jj2\pi\left(k-\frac{K+1}{2}\right)\Delta f\tau^{\rm A}_{u,t,\ell}}
 e^{\jj2\pi\nu^{\rm A}_{u,t,\ell}tT_{\mathrm s}},
\end{aligned}
\label{eq:channel}
\end{equation} 
where $\mathbf h^{\rm A}_{u,t,k}\in \mathbb{C}^{M\times1}$ and $A\in\{\ul,\dl\}$ indicates the UL or DL. Accordingly, $f^{\rm A}$ denotes the corresponding carrier frequency. The number of channel paths is denoted by $L$.
The parameter $\beta^{\rm A}_{u,t,\ell}$ represents the complex path gain of the $\ell$-th path for user $u$ at OFDM symbol $t$. The parameters $\tau^{\rm A}_{u,t,\ell}$, $\phi^{\rm A}_{u,t,\ell}$, $\theta^{\rm A}_{u,t,\ell}$, and $\nu^{\rm A}_{u,t,\ell}$ denote the corresponding delay, azimuth angle, zenith angle, and Doppler shift, respectively. 
The steering vector is given by
\begin{equation}
\mathbf a^{\rm A}(\phi_{u,t,\ell},\theta_{u,t,\ell},f^{\rm A})
=\mathbf a_{\mathrm h}^{\rm A}(\phi_{u,t,\ell},\theta_{u,t,\ell},f^{\rm A}) \otimes \mathbf a_{\mathrm v}^{\rm A}(\theta_{u,t,\ell},f^{\rm A}),
\label{eq:steering_vector}
\end{equation}
where 
\begin{equation}
\mathbf a_{\mathrm h}^{\rm A}(\phi_{u,t,\ell},\theta_{u,t,\ell},f^{\rm A})
=
\begin{bmatrix}
1\\
e^{\mathrm j2\pi\frac{ d_{\mathrm h}f^{\rm A}}{c}
\sin\theta^{\rm A}_{u,t,\ell}\cos\phi^{\rm A}_{u,t,\ell}}
\\
\vdots
\\
e^{\mathrm j2\pi\frac{ d_{\mathrm h}f^{\rm A}}{c}(M_h-1)
\sin\theta^{\rm A}_{u,t,\ell}\cos\phi^{\rm A}_{u,t,\ell}}
\end{bmatrix},
\label{eq:horizontal_steering}
\end{equation}

\begin{equation}
\mathbf a_{\mathrm v}^{\rm A}(\theta_{u,t,\ell},f^{\rm A})
=
\begin{bmatrix}
1\\
e^{\mathrm j2\pi\frac{ d_{\mathrm v}f^{\rm A}}{c}
\cos\theta^{\rm A}_{u,t,\ell}}
\\
\vdots
\\
e^{\mathrm j2\pi \frac{d_{\mathrm v}f^{\rm A}}{c}(M_v-1)
\cos\theta^{\rm A}_{u,t,\ell}}
\end{bmatrix},
\label{eq:vertical_steering}
\end{equation}
and $d_{\mathrm h}$, $d_{\mathrm v}$ are the spacings between the antennas in the horizontal and vertical directions, respectively. $M_h$ and $M_v$ are the numbers of horizontal and vertical antennas at the BS, and $M=M_h\times M_v$. $c$ is the speed of light.
The frequency-domain channel matrix at symbol $t$ is $ \mathbf H^{\rm A}_{u,t}=\left[\mathbf h^{\rm A}_{u,t,1},\ldots,\mathbf h^{\rm A}_{u,t,K}\right] \in\mathbb C^{M\times K}$. 
The channel parameters are assumed to be the same across all subcarriers within each OFDM symbol.

\subsection{UL and DL Transmission Phases}
Following the CSI feedback system design in \cite{9789200}, the UL and DL transmission consist of four phases, which are UL sounding, DL training, UL feedback and DL CSI prediction, as shown in Fig. \ref{fig:system}. Here, we assume that the processing delay between consecutive phases is negligible.
During UL sounding, each UE transmits a SRS to the BS, as specified in the 5G New Radio (NR) standard \cite{3gpp38211}.
The BS then estimates the UL channel parameters at $t_0=0$ and broadcasts pilots to all UEs during DL training over $t_1\in\{1,\ldots,T_1\}$. The UEs transmit their feedback during the UL feedback phase over $t_2\in\{T_1+1,\ldots,T_1+T_2\}$. Finally, the BS predicts the future DL channel for $t_3\in\{T_1+T_2+1,\ldots,T_1+T_2+T_3\}$. The detailed processes of these four phases are described in the subsequent paragraphs.

\paragraph{UL sounding}
At $t_0$, the BS first estimates the UL channel parameters from the SRS transmitted by each UE, resulting in $\beta^{\ul}_{u,t_0,\ell},\tau^{\ul}_{u,t_0,\ell},\phi^{\ul}_{u,t_0,\ell},\theta^{\ul}_{u,t_0,\ell}$, and $\nu^{\ul}_{u,t_0,\ell}$. Among these parameters, the path gain is nonreciprocal, i.e., $\beta^{\ul}_{u,t_0,\ell}\neq\beta^{\dl}_{u,t_0,\ell}$. In contrast, the delays and angles are reciprocal, such that $\tau^{\ul}_{u,t_0,\ell}\approx\tau^{\dl}_{u,t_0,\ell}$, $\phi^{\ul}_{u,t_0,\ell}\approx\phi^{\dl}_{u,t_0,\ell}$, and $\theta^{\ul}_{u,t_0,\ell}\approx\theta^{\dl}_{u,t_0,\ell}$ \cite{9322570}.
Since the Doppler shift depends on the carrier frequency, the UL and DL Doppler shifts satisfy $\nu^{\dl}_{u,t_0,\ell}\approx(f_{\dl}/f_{\ul})\nu^{\ul}_{u,t_0,\ell}$~\cite{9322570}. Here, we assume the UL channel parameters are perfectly estimated and focus on the UL feedback and DL CSI prediction phases.

\paragraph{DL training}
For $t_1\in \{1,\dots,T_1\}$, the BS broadcasts orthogonal pilots to all UEs. The observation at UE $u$ is
\begin{equation}
y_{u,t_1,k}=\mathbf x_{t_1,k}^{\top}\mathbf h^{\dl}_{u,t_1,k}
+w_{u,t_1,k}, 
\label{eq:observation}
\end{equation}
where $\mathbf x_{t_1,k}\in\mathbb C^{M\times1}$ is the orthogonal pilot and $w_{u,t_1,k}\sim\CN(0,\sigma^2)$ is additive white Gaussian noise (AWGN) with zero mean and variance $\sigma^2$. We define $\mathbf Y_u^{T_1}\in\mathbb C^{T_1\times K}$, which collects the observations over $T_1$ OFDM symbols and $K$ subcarriers at UE $u$.  

\paragraph{UL feedback}
For $t_2\in\{T_1+1,\dots,T_1+T_2\}$, the $u$-th UE first compresses its pilot observation $\mathbf Y_u^{T_1}$ into the feedback symbol $\mathbf s_u\in\mathbb C^{T_2\times1}$ using an encoder. The compression ratio is defined as $\rho=T_2/(T_1K)$. The UE then transmits $\mathbf s_u$ to the BS.  
Following \cite{csinet}, we assume that the continuous-valued encoder output $\mathbf s_u$ is directly available at the BS decoder without quantization or transmission distortion.  

\paragraph{DL CSI prediction}
In the last phase, for $t_3\in\{T_1+T_2+1,\ldots,T_1+T_2+T_3\}$, the BS uses the feedback symbols $\mathbf s_u$ to predict the future DL channel parameters $\beta^{\dl}_{u,t_3,\ell},\tau^{\dl}_{u,t_3,\ell},\phi^{\dl}_{u,t_3,\ell},\theta^{\dl}_{u,t_3,\ell}$, and $\nu^{\dl}_{u,t_3,\ell}$ for each UE $u$ and path $\ell$ using a decoder.
The predicted parameters are then used to synthesize the corresponding DL channel matrix $\mathbf H^{\dl}_{u,t_3}\in\mathbb C^{M\times K}$ according to \eqref{eq:channel}.

\begin{figure}[!t]
\centering
\includegraphics[width=0.48\textwidth]{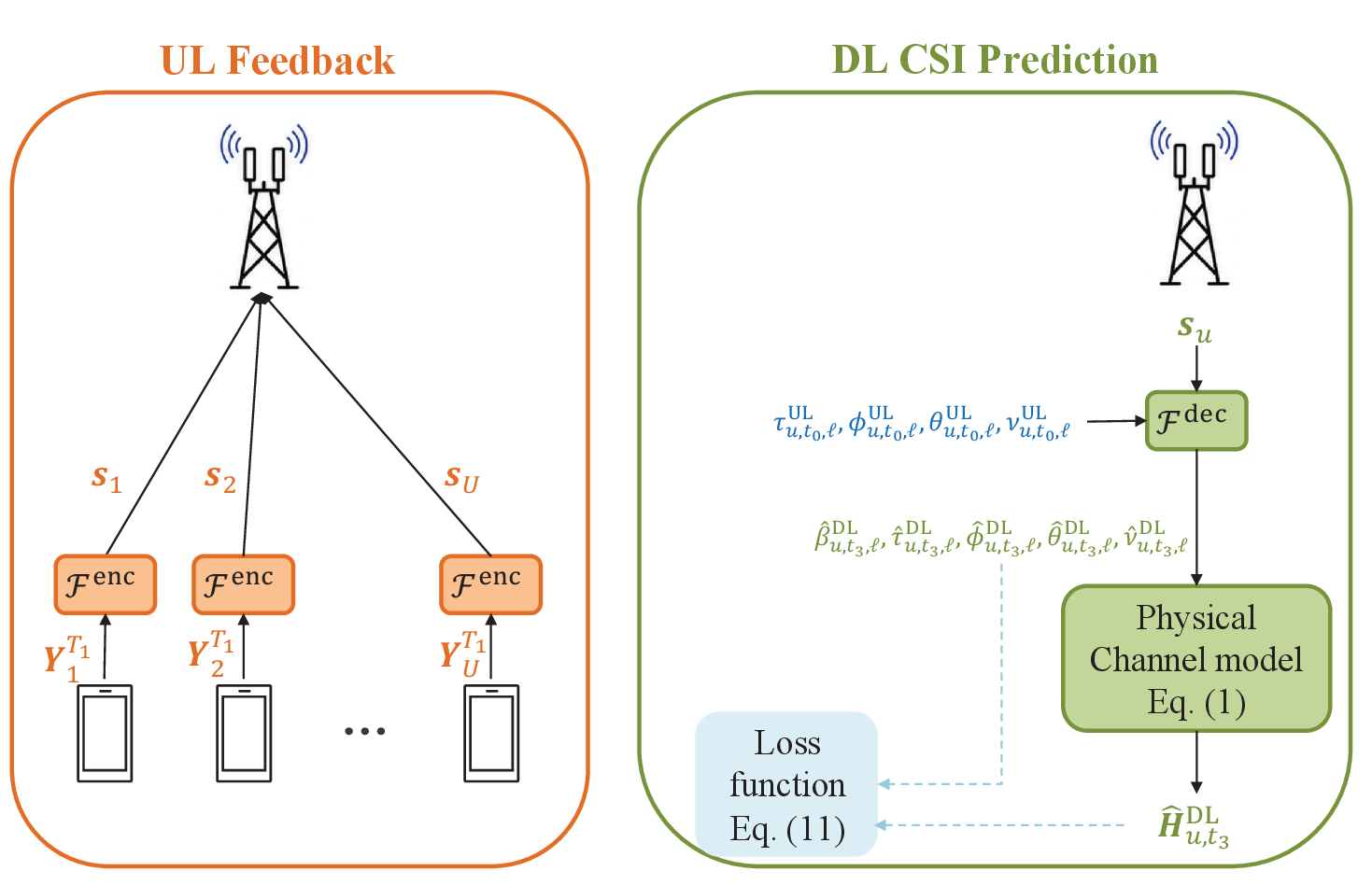}
\caption{Proposed architecture.}
\label{fig:nn}
\vspace{-0.2cm}
\end{figure}

\section{Proposed UL-Assisted Parametric CSI Feedback and Prediction}
The proposed scheme focuses on the UL feedback and DL CSI prediction phases, as shown in Fig. \ref{fig:nn}. During UL feedback, the proposed encoder directly compresses the pilot observations acquired during DL training into feedback symbols without explicit channel estimation at the UE. During DL CSI prediction at the BS, the decoder predicts a set of low-dimensional channel parameters and synthesizes the channel matrix based on these parameters instead of directly outputting a high-dimensional matrix.
In the following, we introduce the proposed UE-side encoder $\fe$ and BS-side decoder $\fd$. Their detailed structures are summarized in Table \ref{tab:architecture}.

\subsection{UE-Side Encoder}
The $u$-th UE compresses its pilot observation into $\mathbf{s}_u$ using the encoder $\fe$, whose parameters are shared by all $U$ UEs. The encoder is implemented using convolutional layers followed by adaptive pooling and a fully connected (FC) layer.
For the $u$-th UE, the received pilot observation ${\mathbf Y}_u^{T_1}$ is first converted to a real-valued representation by stacking its real and imaginary parts along an additional dimension as $\bar{\mathbf Y}_u^{T_1}
=\!\left(\operatorname{Re}\{\mathbf Y_u^{T_1}\},\operatorname{Im}\{\mathbf Y_u^{T_1}\}\right)\in\mathbb R^{2\times T_1\times K}$.
It is then compressed into a real-valued feedback symbol as
\begin{equation}
\bar{\mathbf s}_u
=\fe\!\left(\bar{\mathbf Y}_u^{T_1}\right).
\label{eq:encoder}
\end{equation}
The resulting vector $\bar{\mathbf s}_u\in\mathbb R^{2T_2\times1}$ is normalized as

\begin{equation}
\label{eq:powernorm} 
\mathbf s_u=\sqrt{T_2}\frac{\bar{\mathbf s}_u}{\|\bar{\mathbf s}_u\|_2}. 
\end{equation}
The normalization ensures the unit-average transmit power for feedback symbols. 
Under the ideal UL feedback assumption introduced in Section II, the BS receives $\mathbf s_u$ for $u\in\{1,\dots,U\}$ without any error.

\subsection{BS-Side Decoder}
Based on the received feedback symbol and the reciprocal UL channel parameters, the BS-side decoder $\fd$ predicts future DL channel parameters independently for each UE. $\fd$ employs FC layers to fuse the feedback representation with the UL parameter prior and predict the DL channel parameters.
Specifically, the reciprocal UL parameters are normalized by their respective maximum values, $\tau_{\max}$ and $\nu_{\max}$, before being fed into the decoder. The angular parameters are represented using their sine and cosine values. Here, $\tau_{\max}$ is the maximum delay for both UL and DL, and $\nu_{\max}$ is the maximum DL Doppler shift.  
The resulting UL parameter vector and matrix are respectively defined as
\begin{equation}
\begin{aligned}
{\mathbf p}^{\ul}_{u,t_0,\ell}
=\big[&\tau^{\ul}_{u,t_0,\ell}/\tau_{\max},
\nu^{\ul}_{u,t_0,\ell}/\nu_{\max},
\sin\phi^{\ul}_{u,t_0,\ell},
\cos\phi^{\ul}_{u,t_0,\ell},\\ 
&\sin\theta^{\ul}_{u,t_0,\ell},
\cos\theta^{\ul}_{u,t_0,\ell}\big]^{\top},\\
{\mathbf P}^{\ul}_{u,t_0}
&=\left[{\mathbf p}^{\ul}_{u,t_0,1},\ldots,
{\mathbf p}^{\ul}_{u,t_0,L}\right].
\end{aligned}
\label{eq:decoder_ul_features}
\end{equation}
where ${\mathbf p}^{\ul}_{u,t_0,\ell}\in\mathbb R^{6\times1}$ and ${\mathbf P}^{\ul}_{u,t_0}\in\mathbb R^{6\times L}$.
The UL path gain $\beta^{\ul}_{u,t_0,\ell}$ is excluded from ${\mathbf p}^{\ul}_{u,t_0,\ell}$, since it is not generally reciprocal across UL and DL carrier frequencies, as discussed in Section II. The UL parameters are further vectorized and concatenated with the feedback $\mathbf{s}_u$ as $\mathbf z_u
=\left[{\mathbf s}_u^{\top},\operatorname{vec}\!\left({\mathbf P}^{\ul}_{u,t_0}\right)^{\top}
\right]^{\top}\in\mathbb R^{(2T_2+6L)\times1}. $
Each $\mathbf z_u$ is independently processed by the decoder, givn as
\begin{equation}
\Delta\mathcal P^{\dl}_{u}
=\fd\!\left(\mathbf z_u\right),
\label{eq:decoder_update}
\end{equation}
where $\Delta\mathcal P^{\dl}_{u}\in\mathbb R^{T_3\times 6\times L}$ contains the parameters over the complete prediction period $T_3$.
We let $\Delta\mathbf p^{\dl}_{u,t_3,\ell}\in\mathbb R^{6\times 1}$ represent the parameter vector of path $\ell$ at $t_3$, given by
\begin{equation}
\begin{aligned}
\Delta\mathbf p^{\dl}_{u,t_3,\ell}
 =&\big[
\hat{\beta}^{\dl,\mathrm R}_{u,t_3,\ell},
\hat{\beta}^{\dl,\mathrm I}_{u,t_3,\ell},
\Delta\tau^{\dl}_{u,t_3,\ell},
\Delta\phi^{\dl}_{u,t_3,\ell},\\
&\Delta\theta^{\dl}_{u,t_3,\ell},
\Delta\nu^{\dl}_{u,t_3,\ell}\big],
\label{eq:decoder_update_vector}
\end{aligned}
\end{equation}
where the first two outputs $\hat{\beta}^{\dl, \mathrm R}_{u,t_3,\ell}$ and $\hat{\beta}^{\dl,\mathrm I}_{u,t_3,\ell}$ represent the real and imaginary parts of the future DL path gain $\hat{\beta}^{\dl}_{u,t_3,\ell}$. 
The remaining outputs describe normalized changes relative to the UL parameters at $t_0$. This residual formulation exploits the UL parameters as prior information and requires the decoder to learn only the UL-DL mismatch and temporal variations. These changes typically have smaller dynamic ranges than the absolute parameters, which facilitates network training.
Accordingly, the predicted delay and Doppler are $\hat\tau^{\dl}_{u,t_3,\ell}
=\tau^{\ul}_{u,t_0,\ell}+\tau_{\max}\tanh(\Delta\tau^{\dl}_{u,t_3,\ell})$ and $ \hat\nu^{\dl}_{u,t_3,\ell}=(f_{\dl}/f_{\ul})\nu^{\ul}_{u,t_0,\ell}+\nu_{\max}\tanh(\Delta\nu^{\dl}_{u,t_3,\ell})$, respectively.  
Similarly, the predicted azimuth and zenith are $\hat\phi^{\dl}_{u,t_3,\ell}
=\phi^{\ul}_{u,t_0,\ell}+\phi_{\max}\tanh(\Delta\phi^{\dl}_{u,t_3,\ell})$ and $\hat\theta^{\dl}_{u,t_3,\ell}
=\theta^{\ul}_{u,t_0,\ell}+\theta_{\max}\tanh(\Delta\theta^{\dl}_{u,t_3,\ell})$, respectively. 
$\phi_{\max}$ and $\theta_{\max}$ are their corresponding maximum angles.
The predicted azimuth angle is wrapped to its principal interval, whereas the predicted zenith angle is clipped to $[0,\pi]$. The $\tanh(\cdot)$ activation bounds every normalized change to $[-1,1]$. 
Finally, substituting the predicted parameters into the physical channel model \eqref{eq:channel} yields the future DL channel matrix $\hat{\mathbf H}^{\dl}_{u,t_3}=\left[\hat{\mathbf h}^{\dl}_{u,t_3,1},\ldots,\hat{\mathbf h}^{\dl}_{u,t_3,K}
\right]\in\mathbb{C}^{M\times K}$.

\subsection{End-to-End Training}
The UE encoder $\fe$ and BS decoder $\fd$ are jointly optimized using the predicted channel matrix loss and parameter loss.
The channel matrix loss directly optimizes the accuracy of the synthesized DL channel, whereas the parameter loss provides explicit supervision for each predicted path parameter. Combining these two losses balances end-to-end channel prediction accuracy and parameter-level fidelity, preventing accurate channel reconstruction from being achieved with physically inaccurate parameters.  
The overall loss is given as 
\begin{equation}
    \mathcal L=\lambda_H\mathcal L_{H} + \lambda_P\mathcal L_{P},
\end{equation}
where $\mathcal L_{H}$ and $\mathcal L_{P}$ denote the predicted channel matrix loss and channel parameter loss, respectively. The coefficients $\lambda_H$ and $\lambda_P$ are their corresponding weights in the loss function.
Specifically, $\mathcal L_{H}$ is calculated over $U$ UEs and $T_3$ OFDM symbols, given by
\begin{equation}
\mathcal L_{H}=\frac{1}{UT_3}\sum_{u=1}^{U}\sum_{t_3=T_1+T_2+1}^{T_1+T_2+T_3}
\frac{\|\hat{\mathbf H}^{\dl}_{u,t_3}-\mathbf H^{\dl}_{u,t_3}\|_2^2}
{\|\mathbf H^{\dl}_{u,t_3}\|_2^2},
\label{eq:channel_loss}
\end{equation} 
and $\mathcal L_{P}$ is computed over $L$ paths, $U$ UEs and $T_3$ OFDM symbols as
\begin{equation}
\mathcal L_{P}=\frac{1}{UT_3L}\sum_{u=1}^{U}\sum_{t_3=T_1+T_2+1}^{T_1+T_2+T_3}\sum_{\ell=1}^{L}
\epsilon_{u,t_3,\ell},
\label{eq:para_loss}
\end{equation} 
where
\begin{equation}
\begin{aligned}
\epsilon_{u,t_3,\ell}
={}&
\left|
\hat{\beta}^{\dl}_{u,t_3,\ell}
-\beta^{\dl}_{u,t_3,\ell}
\right|^2
\\
&+
\left(
\frac{
\hat{\tau}^{\dl}_{u,t_3,\ell}
-\tau^{\dl}_{u,t_3,\ell}
}{
\tau_{\max}
}
\right)^2
+
\left(
\frac{
\hat{\nu}^{\dl}_{u,t_3,\ell}
-\nu^{\dl}_{u,t_3,\ell}
}{
\nu^{\dl}_{\max}
}
\right)^2
\\
&+
\left(
\frac{
\hat{\phi}^{\dl}_{u,t_3,\ell}
-\phi^{\dl}_{u,t_3,\ell}
}{
\phi_{\max}
}
\right)^2
+
\left(
\frac{
\hat{\theta}^{\dl}_{u,t_3,\ell}
-\theta^{\dl}_{u,t_3,\ell}
}{
\theta_{\max}
}
\right)^2 .
\end{aligned}
\label{eq:parameter_error}
\end{equation}
The NN parameters in $\fe$ and $\fd$ are jointly optimized using the Adam optimizer.
%\red{Why use this type of loss? the reasons contain trade-off between...., which is not expressed here}
 \begin{table*}[!t]
\centering
\caption{NN architectures of $\fe$ and $\fd$}
\label{tab:architecture}
\small
\setlength{\tabcolsep}{8pt}
\renewcommand{\arraystretch}{1.25}
\begin{tabular*}{\textwidth}{@{\extracolsep{\fill}}clll@{}}
\toprule
Module & Layer & Input shape & Output shape \\
\midrule
\multirow{5}{*}{$\fe$: UE encoder}
& Conv2D ($2\!\rightarrow\!32$, $3\!\times\!3$, $\mathrm {stride}\!=\!1$) + GELU & $2\times T_1\times K$ & $32\times T_1\times K$ \\
& Conv2D ($32\!\rightarrow\!64$, $3\!\times\!3$, $\mathrm {stride}\!=\!2$) + GELU & $32\times T_1\times K$ & $64\times \frac{T_1}{2}\times \frac{K}{2}$ \\
& Conv2D ($64\!\rightarrow\!64$, $3\!\times\!3$, $\mathrm {stride}\!=\!2$) + GELU & $64\times \frac{T_1}{2}\times \frac{K}{2}$ & $64\times \frac{T_1}{4}\times \frac{K}{4}$ \\
& Adaptive average pooling ($4\times4$) & $64\times \frac{T_1}{4}\times \frac{K}{4}$ & $64\times4\times4$ \\
& Flatten + FC  & $64\times4\times4$ & $2T_2$ \\ 
\midrule
\multirow{3}{*}{$\fd$: BS decoder}
& FC  + GELU & $2T_2+6L$ & $F$ \\
& FC  + GELU & $F$ & $F$ \\
& FC  & $F$ & $T_3\times 6\times L$ \\
\bottomrule
\vspace{-1cm}
\end{tabular*} 
\end{table*}

\section{Simulation Results}
\subsection{System Configurations}
We generate the dataset using Sionna RT \cite{sionnaRT} with $U=4$ UEs, $M=4$ BS antennas, and $K=32$ subcarriers. The subcarrier spacing is 15 kHz, while the UL and DL carrier frequencies are $f^{\ul}=3.4$ GHz and $f^{\dl}=3.5$ GHz, respectively. We consider $L=4$ channel paths for each UE and set the UE speed to 108 km/h. The numbers of OFDM symbols are $T_1=16$, $T_2=8$ and $T_3=32$ for DL training, UL feedback and DL CSI prediction, respectively. Accordingly, the compression ratio is $\rho=1/64$.
We set the maximum delay and Doppler shift to $\tau_{\rm max}=220$ ns and $\nu^{\dl}_{\rm max}=350$ Hz, respectively, and set the maximum azimuth and zenith angles to $\phi_{\rm max}=\theta_{\rm max}=\pi$.
As each UE moves along its trajectory, the changing propagation geometry produces temporal variations in the path gains, delays, angles, and Doppler shifts.
The UL channel parameters at $t_0$ are assumed to be perfectly known, and the DL training uses all subcarriers for pilot symbols.

The NN architectures of $\fe$ and $\fd$ are summarized in Table~\ref{tab:architecture}, including their layer settings and input and output dimensions. The decoder hidden layer size is set to $F=32$. The loss weights are $\lambda_H=0.8$ and $\lambda_P=0.2$.
We collect 50,000 data samples from Sionna RT and split them into 40,000, 5,000 and 5,000 for training, validation and testing, respectively. All models are trained for 100 epochs with a batch size of 256 and a learning rate of $5\times10^{-4}$. During training, each batch of samples is assigned an SNR value selected from $\{0,5,10,15,20,25\}$ dB.

\subsection{Baseline Definitions} 
We consider three baselines with the same budget of $T_2$ feedback symbols and the same power normalization as the proposed scheme. They follow the concepts of CsiNet \cite{csinet} and PFNet \cite{GUO202483}. For a fair comparison, they use the same encoder and decoder layers and the same hidden layer size as the proposed scheme, with their input and output dimensions adjusted according to their respective tasks.

\begin{itemize}
\item \emph{Baseline 1:} Similar to CsiNet \cite{csinet}, the UE first applies a least-squares (LS) channel estimator to obtain the DL channel matrix over the $T_1$ DL training period. Its encoder maps the estimated channel matrix into feedback symbols. The BS decoder directly predicts the future DL channel matrix from the received feedback without using the reciprocal UL channel parameters or the physical channel model. 

\item \emph{Baseline 2:} Similar to PFNet~\cite{GUO202483}, the UE directly compresses the pilot observations obtained during DL training into feedback symbols without explicit channel estimation. The BS decoder, which is the same as that of Baseline 1, directly predicts the future DL channel matrix from the received feedback without using the reciprocal UL channel parameters or the physical channel model.

\item \emph{Baseline 3:} Similar to the proposed scheme, the UE directly compresses the pilot observations into feedback symbols. The BS decoder combines the received feedback symbols with the reciprocal UL channel parameters but predicts only the future DL channel path gains. The future DL channel matrix is then synthesized using the predicted gains while reusing the outdated UL delays, angles, and Doppler shifts obtained at $t_0$. 

\end{itemize}
%\red{Maybe one more baseline that has no UL SRS, to observe how much influence the UL SRS can bring}
All baselines use the same datasets, feedback budget, and training configuration as the proposed scheme.

\begin{figure}[t]
\centering
\includegraphics[width=0.8\columnwidth]{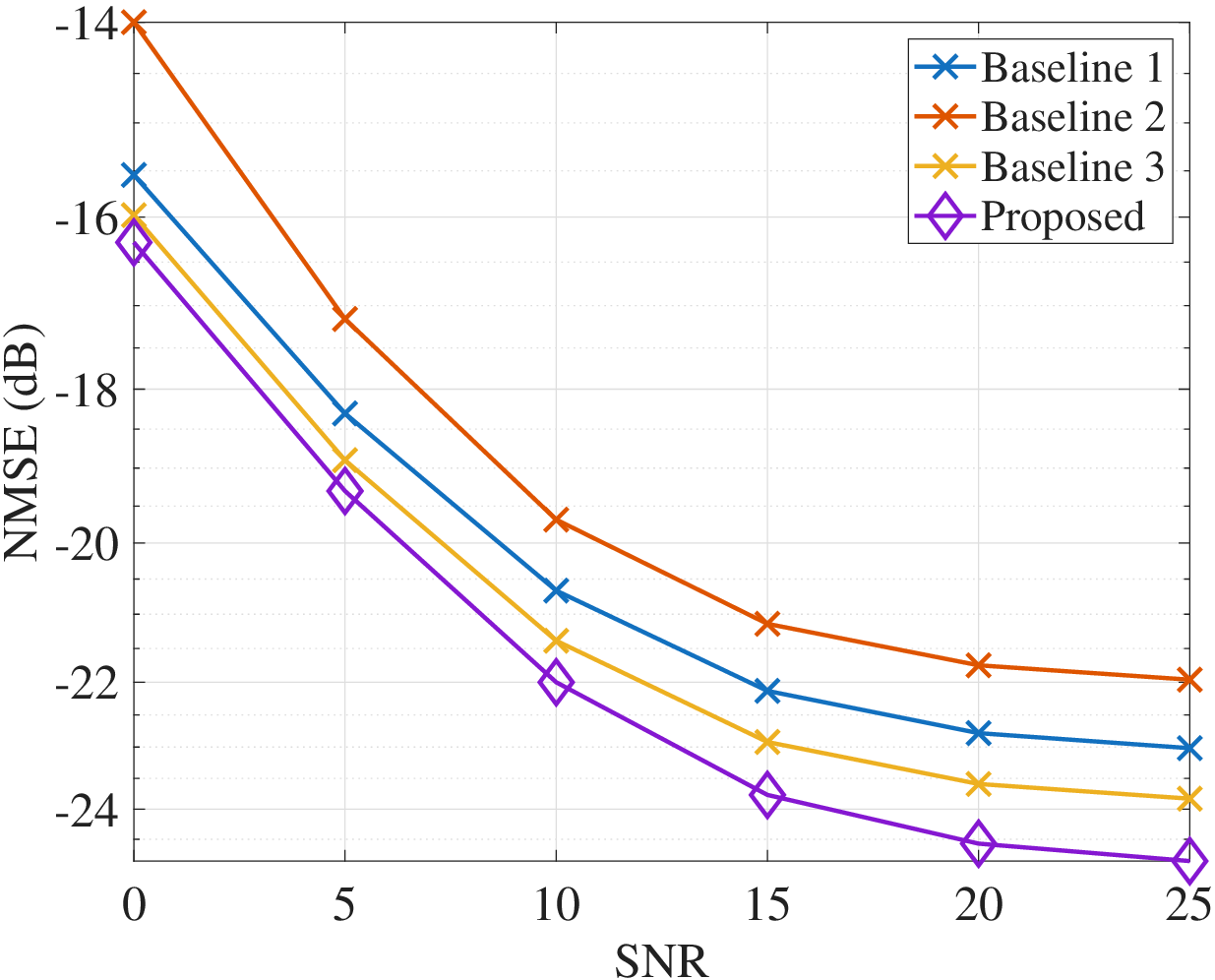}
\caption{NMSE performance for different DL SNRs.}
\label{fig:snr} 
\vspace{-0.2cm}
\end{figure}

\subsection{Channel Prediction Performance} 

We first evaluate the channel prediction performance of the proposed scheme over different DL pilot SNRs. As shown in Fig. \ref{fig:snr}, the proposed scheme consistently outperforms all evaluated baselines. In particular, it outperforms Baseline 3 because both schemes predict the future DL path gains, but only the proposed decoder predicts the changes in the delays, angles, and Doppler shifts relative to the UL parameters at $t_0$. Therefore, the proposed scheme can better compensate for channel variations.
Baseline 2 performs worse than Baseline 1 because it has to implicitly extract the channel information from noisy pilot observations, whereas Baseline 1 encodes an explicitly estimated channel matrix. This observation is consistent with the results reported in \cite{GUO202483}. Baseline 3 also outperforms Baselines 1 and 2 because it exploits the UL channel parameters and the physical channel model. Consequently, its decoder predicts only low-dimensional path gains instead of a high-dimensional channel matrix. 
 
\begin{figure}[t]
\centering
\includegraphics[width=0.8\columnwidth]{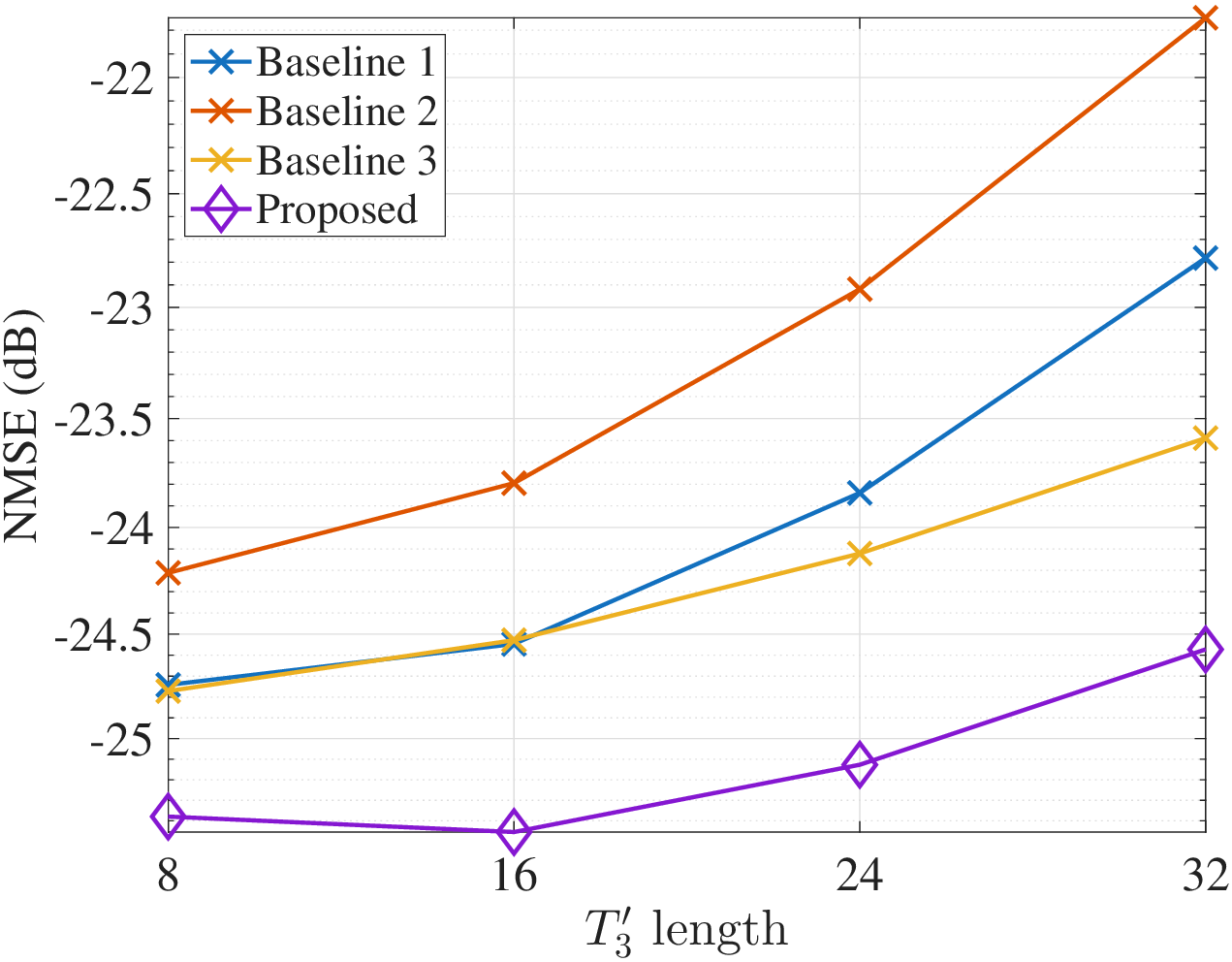}
\caption{NMSE performance for different $T_3'$ lengths at $SNR=20$ dB.}
\vspace{-0.2cm}
\label{fig:p3} 
\end{figure}

We then evaluate the channel prediction performance for different prediction lengths at a fixed DL pilot SNR of 20 dB. All the schemes are trained once with a maximum prediction length of $T_3=32$. For each evaluation length $T_3'\in\{8,16,24,32\}$, the reported NMSE is averaged over the first $T_3'$ predicted OFDM symbols.
As shown in Fig. \ref{fig:p3}, the proposed scheme achieves the lowest NMSE for all the evaluated values of $T_3'$. 
Its NMSE decreases slightly up to $T_3'=16$ but increases for larger $T_3'$. This is because the network is jointly trained over the entire $T_3=32$ prediction interval, and it minimizes the average NMSE across all predicted symbols rather than the error at each individual time instant. 
Moreover, the consistent outperformance over Baseline 3 demonstrates the benefit of predicting channel parameter changes instead of directly reusing the outdated parameters. Baseline 3 outperforms Baselines 1 and 2 because it exploits the UL channel parameters and the physical channel model. In contrast, Baselines 1 and 2 directly predict the high-dimensional channel matrix and therefore exhibit greater performance degradation as the prediction length increases.

\subsection{Complexity Analysis}
We evaluate the model complexity in terms of the total number of trainable parameters, as summarized in Table~\ref{tab:complexity}. 
The number of UE encoder parameters is identical for all schemes because they use the same convolutional layers, while the adaptive $4\times4$ pooling produces a fixed output size independent of the encoder input dimensions. 
Nevertheless, an equal number of parameters does not imply equal computational cost. Baseline~1 processes a $T_1\times KM$ estimated channel matrix instead of a $T_1\times K$ pilot observation and additionally requires explicit LS channel estimation. Therefore, it has a higher UE-side computational cost than Baseline~2, although the two baselines have the same number of trainable parameters. In contrast, the proposed scheme uses approximately $70\%$ fewer trainable parameters than Baselines~1 and 2 because it predicts low-dimensional parameters instead of a high-dimensional channel matrix. Baseline~3 has the fewest trainable parameters because it predicts only the channel path gains.  

\begin{table}[t]
\centering
\caption{Trainable parameters}
\label{tab:complexity}
\scriptsize
\setlength{\tabcolsep}{1.2pt}
\begin{tabularx}{\columnwidth}{@{}>{\raggedright\arraybackslash}Xrrrr@{}}
\toprule
Module & Proposed & Baseline 1 & Baseline 2 & Baseline 3 \\
\midrule
UE encoder  & $72,432$ & $72,432$ & $72,432$ & $72,432$ \\
BS decoder  & $27,712$ & $271,936$ & $271,936$ & $10,816$ \\
Total & $100,144$ & $344,368$ & $344,368$ & $83,248$ \\  
\bottomrule
\vspace{-1cm}
\end{tabularx}
\end{table}
 
\section{Conclusion}
This paper proposes a UL-assisted parametric CSI feedback scheme for channel prediction in FDD systems. Each UE directly maps its DL pilot observations to feedback symbols, thereby avoiding explicit UE-side channel estimation. The BS combines these feedback symbols with the UL parameters to predict future DL channel parameters and synthesize the DL channel matrix using an explicit physical channel model. Simulation results on Sionna RT dataset show that the proposed scheme achieves lower NMSE than all evaluated baselines across all DL pilot SNRs and various prediction lengths.

\bibliographystyle{IEEEtran}
\bibliography{main}
\end{document}